\documentclass[a4paper,12pt]{article}
\usepackage{tikz}
\usepackage{tikz-3dplot}
\usetikzlibrary{decorations.pathreplacing,shapes,calc,positioning}
\tdplotsetmaincoords{60}{60}
\usepackage{amsmath}
\usepackage{graphicx,subcaption}
\usepackage{amssymb}
\usepackage{adjustbox}
\usepackage{multirow}
\usepackage{color}
\usepackage{float}
\usepackage{soul}
\usepackage{booktabs}
\usepackage{natbib}
\usepackage{upgreek}
\usepackage{bm}
\usepackage{url}

\title{Comparing probabilistic predictive models applied to football}
\author{Marcio Alves Diniz, Rafael Izbicki, \\Danilo Lopes and Luis Ernesto Salasar}
\date{}

\begin{document}

\maketitle

\abstract{We propose two Bayesian multinomial-Dirichlet models to predict the final outcome of football (soccer) matches and compare them to three well-known models regarding their predictive power. All the models predicted the full-time results of 1710 matches of the first division of the Brazilian football championship and the comparison used three proper scoring rules, the proportion of errors and a calibration assessment. We also provide a goodness of fit measure.
Our results show that multinomial-Dirichlet models are not only competitive
with standard approaches, but they are also well calibrated and present reasonable goodness of fit.

}


\section{Introduction}
	
	

Several models
for football (soccer) prediction exist (see, e.g.,  \cite{Owen2011,Koopman2015, Volf2009, Titman2015} and references therein).
In this work, we  (i) propose two novel Bayesian
multinomial-Dirichlet models that consider only the number of matches won, drawn or lost by each team as inputs, and (ii)  compare such models with two benchmark models, whose predictions for matches of the Brazilian national championships are published on Internet websites---see \cite{reference13} and \cite{reference14}. 
Such models are widely consulted by football fans and consider multiple covariates as inputs.
 As a baseline, we also
make comparisons with  an extension of the Bradley-Terry model \citep{davidson1970}.

Brazilian football championships are disputed by 20~teams that play against each other twice (home and away) and the team with more points after all matches are played is declared champion.
	Therefore, 380 matches are played per championship, 190 in each half.
	The last four teams are relegated to a minor division and the first four play Copa Libertadores (South America champions league).
	Our analysis comprised the championships from 2006 to 2014, because it was only in 2006 that this form of dispute was implemented in the Brazilian national championships.

Our comparisons were  made
using  1710 matches of the first division of the Brazilian football championship. Several  standard metrics (scoring rules) were used for ranking the models, as well as
 other criteria such as the proportion of matches that were ``incorrectly'' predicted by each model and a measure of calibration.

There are several ways to score or classify predictions of categorical events that assume one result out of a discrete set of mutually exclusive possible outcomes, like football matches.
See \cite{constantinou} for a brief survey of such measures applied to football.
We decided to score the predictions for each match in terms of their distances from the truth, \emph{i.e.}, the verified event, once it has occurred, and chose the most used distances in the literature: Brier \citep{brier1950}, logarithmic and spherical.
	
This paper is organized as follows.
Section \ref{sec::experimental} describes the studied models, 
Section \ref{sec::results} reports the predictive performance of the models and a goodness of fit measure.
In Section \ref{sec::remarks} we discuss the results and close with proposals for future research.
The Appendix briefly describes the scoring rules and the criteria used in this work to classify the models.
		
\section{The Models: Theoretical Background}
\label{sec::experimental}
	
In the statistical literature, the final outcome of football matches is usually predicted by modeling either the number of goals \citep{Maher82, Dixon97, Lee97, Karlis2003} or the categorical final outcome itself (win, draw or loss of the home team)~\citep{Forrest2000, Koning2000, Brillinger2008, Brillinger2009}. 
For a discussion of these two approaches see \cite{Goddard2005}.  In this work, we consider two benchmark models that follow the first approach:  ``Arruda'' (\url{www.chancedegol.com.br}) and ``Lee'' (\url{www.previsaoesportiva.com.br}); the Bradley-Terry model and the models proposed by us follow the second approach.
The predictions of the two benchmark models were published before each matchday.
We use this section to describe these models in some detail.

	\subsection{Benchmark Models}
	\label{sec::Benchmark}
	
	The benchmark models, Arruda \citep{arruda2000} and Lee \citep{Lee97}, assume that the number of goals $(Y_1, Y_2)$ scored respectively by teams $A$ (home team) and $B$ (away
	team) has a bivariate Poisson distribution \citep{Holgate64} with parameters $(\uplambda_1, \uplambda_2, \uplambda_3)$,
	which has probability mass function given by
	
	\begin{equation*}
	P(Y_1 = y_1, Y_2 = y_2 | \uplambda_1, \uplambda_2, \uplambda_3) =
	\exp\{-(\uplambda_1 + \uplambda_2 + \uplambda_3)\}
	\sum_{k = 0}^{\min(y_1, y_2)} \dfrac{\uplambda_1^{y_1 - k} \uplambda_2^{y_2 - k} \uplambda_3^k}{(y_1-k)!(y_2 - k)!k!}, \label{eq::pois.biv}
	\end{equation*}
	for $\uplambda_1, \uplambda_2 > 0$ and $\uplambda_3 \geq 0$.

	Both marginal distributions of $(Y_1, Y_2)$ have Poisson
	distributions with dependence parameter $\uplambda_3 \geq 0$. If
	$\uplambda_3 = 0$ the marginal distributions are independent, while if
	$\uplambda_3 > 0$ the marginal distributions are positively
	correlated. 
    While the Lee model assumes that $\lambda_3=0$, the Arruda model does not.
    Because of its flexibility, \cite{Karlis2003} argue
	that this distribution is a plausible choice for modeling dependence
	of scores in sports competitions.

	Similar to \cite{Karlis2003} and \cite{Lee97}, both benchmark models
	adopt the following log-linear link functions
	\begin{align*}
	\log(\uplambda_1) &= \upmu + \text{ATT}_A - \text{DEF}_B + \upgamma, \\
	\log(\uplambda_2) &= \upmu + \text{ATT}_B - \text{DEF}_A,
	\end{align*}
	where $\upmu$ is a parameter representing the average number of goals
	in a match, $\text{ATT}_k$ is the offensive strength of team $k$,
	$\text{DEF}_k$ is the defensive strength of team $k$ and $\upgamma$ is
	the home advantage parameter, $k = A, B$. For both the Arruda and Lee models, it is usual to assume the following identifiability constraint	
	\begin{equation*}
	\sum_{t = 1}^T \text{ATT}_t  = 0, \quad \sum_{t = 1}^T \text{DEF}_t = 0,\\
	\end{equation*}
	where $T$ is the number of teams of the analyzed championship.
	
    The predictions of an upcoming matchday are obtained by fitting the model to all relevant previous observed data and then summing up the probabilities of all scores relevant to the win, draw and loss outcomes. 
    We should remark, however, that the Arruda model uses results of the previous twelve months to predict future matches, but we have no information about how this is done.
    On the other hand, the Lee model uses only information of the current championship.
	

    \subsection{Bradley-Terry model}
\label{sec::BTmodel}

The Bradley-Terry paired comparison model \citep{BradleyTerry1952} was primarily developed for modeling the subjective preference of a set of objects when compared in pairs by one or more judges. 
Applications of this model include studies of preference and choice behaviour, but also the ranking of competitors and the prediction of outcomes in sports, such as chess, tennis and soccer. We consider an extension of the Bradley-Terry model, the Davidson tie model with multiplicative order effects \citep{davidson1977extending}, that allows us to account for both ties and home-field advantage:

\begin{align}
p^W_{ij} &= P(\mbox{Home team } i \mbox{ beats visiting team } j) = \frac{\gamma \pi_i}{\gamma \pi_i + \pi_j + \nu \sqrt{\pi_i\pi_j}} \nonumber\\
p^D_{ij} &= P(\mbox{Home team } i \mbox{ ties with visiting team } j) = \frac{\nu \sqrt{\pi_i\pi_j}}{\gamma \pi_i + \pi_j + \nu \sqrt{\pi_i\pi_j}} \nonumber\\
p^L_{ij} &= P(\mbox{Home team } i \mbox{ loses to visiting team } j) = 1 - p^W_{ij} - p^D_{ij},
\label{eq:BTmodel}
\end{align}
where $\gamma > 0$ is the home advantage parameter, $\nu > 0$ is the parameter that accomodates for draws and $\pi_i$ is the worth parameter, the relative ability of team $i$. To ensure identifiability, it is commonly assumed that $\pi_i \geq 0$ and $\sum \pi_i = 1$.

Maximum likelihood estimation is performed by numerically maximizing the reparameterized log-likelihood function corresponding to an unrestricted lower dimension parameter space. For every upcoming second-half matchday, MLEs are recalculated using the outcomes of all the previous matches (including first and second-half matches) and then plugged in (\ref{eq:BTmodel}) in order to obtain predictions for the new matchday. For a study on the conditions for the existence and uniqueness of the MLE and the penalized MLE for different extensions of the Bradley-Terry model, see \cite{Yan2016}.

	\subsection{Multinomial-Dirichlet}
	\label{sec::Mn_Dir}
	
	Now we explain the Bayesian approach developed in this work to calculate the prediction probabilities of an upcoming match of a given team $A$ based on its past performance, \emph{i.e.}, the number of matches it has won, drawn and lost.
	
	Let us consider the outcome of a given match of team $A$ as a categorical random quantity $X$ that may assume only the values $1$ (if team $A$ wins), $2$ (if a draw occurs), $3$ (if team $A$ loses).
	Denoting by $\uptheta_1, \uptheta_2$ and $\uptheta_3$ (where $\uptheta_3 = 1-\uptheta_1 - \uptheta_2$), the probabilities of win, draw and loss, respectively, the probability mass function of $X$ is
	\[
	P(X=x | \boldsymbol{\uptheta}) = \uptheta_1^{\mathbb{I}_{\{1\}}(x)}
	\uptheta_2^{\mathbb{I}_{\{2\}}(x)}(1 - \uptheta_1 -
	\uptheta_2)^{{\mathbb{I}_{\{3\}}}(x)}, \qquad x \in \mathcal{X},
	\]
	
	\noindent
	where $\mathcal{X}=\{1,2,3\}$ is the support of $X$,
	$\mathbb{I}_{\{i\}}(x)$ is the indicator function that assumes the
	value 1 if $x$ equals $i$ and 0 otherwise, and $\boldsymbol{\uptheta}
	= (\uptheta_1, \uptheta_2)$ belongs to $\boldsymbol{\Theta} =
	\{(\uptheta_1,\uptheta_2)\in [0,1]^2: \uptheta_1+\uptheta_2 \leq 1 \}$, the~2-simplex.
	
	Assuming that the outcomes from $n$ matches of team $A$, given $\boldsymbol{\uptheta}$, are i.i.d. quantities with the above categorical distribution, and denoting by $M_1$, $M_2$ and $M_3$ the number of matches won, drawn or lost by team $A$, the random vector $(M_1, M_2, M_3)$ has Multinomial (indeed, trinomial) distribution with parameters $n$ and $\boldsymbol{\uptheta}$ given by

	\[
	P(M_1=n_1,M_2=n_2,M_3=n_3| n, \boldsymbol{\uptheta})=
	{n \choose n_1, n_2, n_3}\uptheta_1^{n_1}\uptheta_2^{n_2}(1-\uptheta_1-\uptheta_2)^{n_3},
	\]
	\noindent
	where $n_1 + n_2 + n_3 = n$.
	
	Our goal is to compute the predictive posterior distribution of the
	upcoming match, $X_{n+1}$, that is,
	$P(X_{n+1}=x|M_1=n_1,M_2=n_2,M_3=n_3)$, $x\in\mathcal{X}$. 
	Suppose that $\boldsymbol{\uptheta}$ has Dirichlet prior distribution
	with parameter $(\upalpha_1,\upalpha_2,\upalpha_3)$, denoted
	$\mathcal{D}(\upalpha_1,\upalpha_2,\upalpha_3)$, with density function
	\[
	\pi(\boldsymbol{\uptheta}|\boldsymbol{\upalpha})=\frac{\Gamma(\upalpha_1+\upalpha_2+\upalpha_3)}{\Gamma(\upalpha_1)\Gamma(\upalpha_2)\Gamma(\upalpha_3)}\uptheta_1^{\upalpha_1-1}\uptheta_2^{\upalpha_2-1}(1-\uptheta_1-\uptheta_2)^{\upalpha_3-1}
	\]
	\noindent for $\upalpha_1$, $\upalpha_2$, $\upalpha_3 > 0$, then the
	posterior distribution of $\boldsymbol{\uptheta}$ is
	$\mathcal{D}(n_1+\upalpha_1,n_2+\upalpha_2,n_3+\upalpha_3)$. Thus, the
	predictive distribution of $X_{n + 1}$ is given by the
	integral
	$$
	P(X_{n + 1} = x | M_1 = n_1, M_2 = n_2, M_3 = n_3) =\\ \int_{\boldsymbol{\uptheta}} P(X_{n
		+ 1} = x | \boldsymbol{\uptheta}) \pi(\boldsymbol{\uptheta} | M_1 = n_1, M_2 = n_2, M_3
	= n_3) d\boldsymbol{\uptheta},
	$$
	which leads to the following probabilities of win, tie and loss:
	\begin{align*}
	P(X_{n+1} = 1 | M_1=n_1,M_2=n_2,M_3=n_3) &=
	\frac{n_1+\upalpha_1}{n+\upalpha_{\bullet}}\\
	& \\
	P(X_{n+1} = 2 | M_1=n_1,M_2=n_2,M_3=n_3) &=
	\frac{n_2+\upalpha_2}{n+\upalpha_{\bullet}} \\
	& \\
	P(X_{n+1} = 3 | M_1=n_1, M_2=n_2, M_3=n_3) &=
	\frac{n_3+\upalpha_3}{n+\upalpha_{\bullet}}
	\end{align*}
	\noindent where $\upalpha_{\bullet} =\upalpha_1+\upalpha_2+\upalpha_3$.
	 In fact, the multinomial-Dirichlet is a classical model used in several applied works and more information about it can be found in \cite{good1965,bernardo1994} and references therein.
	
In the next subsections, \ref{sec::Mn_Dir1} and \ref{sec::Mn_Dir2}, we propose two multinomial-Dirichlet models (Mn-Dir1 and Mn-Dir2) to predict the second-half matches of the championships given all the previous observed results of the same championship. The first-half results are used to build the prior distribution and the second-half results are applied to assess the predictive power of the models. The homepage that publishes the Arruda model also provides predictions for the first-half matches (using results of the previous twelve months), but we have no specific information about how this is done. Therefore, at the beginning of the championships, we may say that the Dirichlet-multinomial models and the Lee model are handicapped when compared to the Arruda model. Trying to compensate this handicap, we compared the models using just the second-half predictions.
	
Before we explain and illustrate the Dirichlet-multinomial models with an example, we make two further remarks. The first one is that we will separately consider home and away games for each team, allowing us to take into account the different performances under these conditions. The second remark is that, using the Dirichlet-multinomial approach, it is possible to predict the result of an upcoming match between teams $A$ (home team) and $B$ (away team) using the past performance of either teams. An analogy can be made to a situation where there exist two observers: one only informed about the matches A played at home and the other only informed about the matches B played away, each one providing distinct predictive distributions. Then, we propose to combine these predictive distributions by applying the so-called \textit{linear opinion pooling} method, firstly proposed by \cite{Stone61}, which consists of taking a weighted average of the predictive distributions. This method is advocated by \cite{McConway81} and \cite{Lehrer83} as the unique rational choice for combining different probability distributions. For a survey on different methods for combining probability distributions we refer to \cite{Genest86}.

    \subsection{Model One: Multinomial-Dirichet $1$}
	\label{sec::Mn_Dir1}
	
	The model $Mn-Dir_1$ is defined as a mixture with equal weights of two Dirichlet distributions: The posterior distributions of teams $A$ and $B$.
	Since teams $A$ and $B$ are the home and away teams, respectively, the two posterior distributions to be mixed are: (i) one considering only the matches $A$ played at home; (ii) another considering only the matches $B$ played away.
	The relevant past performance of teams $A$ and $B$ will be summarized, respectively, by the count vectors ${\bf h} = (h_1, h_2, h_3)$ (team $A$ at home) and ${\bf a} = (a_1, a_2, a_3)$ (team $B$ away), representing the numbers of matches won, drawn and~lost,~respectively.

Predictions are carried out by using a Bayes information updating mechanism. First, we use full-time results from the first-half matches as historical data for construction of the Dirichlet prior: we assign an uniform prior on $\boldsymbol{\uptheta}$ over the 2-simplex, \emph{i.e.}, $\mathcal{D}(1, 1, 1)$, and combine this prior with the data on the first half of the championship, obtaining a posterior Dirichlet distribution through conjugacy that represents the information about $\boldsymbol{\uptheta}$ up to the first half. Then, the posterior of the first half becomes the prior for the second half, which, for every matchday in the second half, will be combined with all the observed second half matches up to that matchday in order to yield posterior predictive distributions. For more on the uniform prior on the simplex, see \cite{good1965} and \cite{agresti2010}.
	
For instance, consider the match Gr\^emio versus Atl\'etico-PR played for matchday 20 of the 2014 championship, at Gr\^emio~stadium. Table \ref{tab:counts} displays the performances of both teams, home and away, after 19 matches. The~relevant vector of counts to be used are ${\bf h}=(h_1,h_2,h_3)=(6,2,1)$ and ${\bf a}=(a_1,a_2,a_3)=(2,3,4)$. Therefore, Gr\^emio has a $\mathcal{D}(7,3,2)$ posterior for matches played at home and Atl\'etico has a $\mathcal{D}(3,4,5)$ posterior for matches played as visitor (recall that both priors were	$\mathcal{D}(1,1,1)$).
	
	\begin{table}[H]
	\small
		\begin{center}
			\begin{tabular}{lccccccccc}
				
				\toprule
				& \multicolumn{3}{c}{\textbf{Home}} & \multicolumn{3}{c}{\textbf{Away}}& \multicolumn{3}{c}{\textbf{Overall}} \\
				\midrule
				Team & W & D & L & W & D & L & W & D & L\\

				Gr\^emio & 6 & 2 & 1 & 3 & 2 & 5 & 9 & 4 & 6\\
				Atl\'etico-PR & 4 & 4 & 2 & 2 & 3 & 4 & 6 & 7 & 6\\
				\bottomrule
			\end{tabular}
			\caption{Gr\^emio and Atl\'etico-PR counts after 19 matchdays (first half).}\label{tab:counts}
		\end{center}
	\end{table}
	
	Thus, considering $X_{n + 1}$ the random outcome of this match with
	respect to the home team (Gr\^{e}mio), the predictive probabilities
	of $X_{n + 1}$ is obtained by equally weighting the two predictive
	distributions,~resulting
	\begin{align*}
	P(X_{n+1}=1|{\bf h}, {\bf a}) &=
	\frac{1}{2}\left(\frac{h_1+\upalpha_1}{h_{\bullet}+\upalpha_{\bullet}}\right)+\frac{1}{2}\left(\frac{a_3+\upalpha_3}{a_{\bullet}+\upalpha_{\bullet}}\right)=0.5
	\\
	P(X_{n+1}=2|{\bf h}, {\bf a}) &=
	\frac{1}{2}\left(\frac{h_2+\upalpha_2}{h_{\bullet}+\upalpha_{\bullet}}\right)+\frac{1}{2}\left(\frac{a_2+\upalpha_2}{a_{\bullet}+\upalpha_{\bullet}}\right)\simeq0.2917, \\
	P(X_{n+1}=3|{\bf h}, {\bf a}) &= \frac{1}{2}
	\left(\frac{h_3+\upalpha_3}{h_{\bullet}+\upalpha_{\bullet}}\right)+\frac{1}{2}\left(\frac{a_1+\upalpha_1}{a_{\bullet}+\upalpha_{\bullet}}\right)\simeq0.2083.
	\end{align*}
	\noindent where $h_{\bullet}=h_1+h_2+h_3$ and $a_{\bullet}=a_1+a_2+a_3$.
	
	\subsection{Model Two: Multinomial-Dirichlet $2$}
	\label{sec::Mn_Dir2}
	
The model $Mn-Dir_2$ is similar to model $Mn-Dir_1$, except that the weights can be different and the chosen prior for $\boldsymbol{\theta}$ is now a symmetric Dirichlet $\mathcal{D}(\alpha,\alpha,\alpha)$, $\alpha > 0$. Thus, the predictive probabilities of $X_{n+1}$ are given by
	\begin{align*}
	P(X_{n+1}=1|{\bf h}, {\bf a}) &=
	w \left(\frac{h_1+\upalpha}{h_{\bullet}+\upalpha_{\bullet}}\right)+(1 - w) \left(\frac{a_3+\upalpha}{a_{\bullet}+\upalpha_{\bullet}}\right) \\
	P(X_{n+1}=2|{\bf h}, {\bf a}) &=
	w \left(\frac{h_2+\upalpha}{h_{\bullet}+\upalpha_{\bullet}}\right)+(1-w) \left(\frac{a_2+\upalpha}{a_{\bullet}+\upalpha_{\bullet}}\right), \\
	P(X_{n+1}=3|{\bf h}, {\bf a}) &= w \left(\frac{h_3+\upalpha}{h_{\bullet}+\upalpha_{\bullet}}\right) + (1-w) \left(\frac{a_1+\upalpha}{a_{\bullet}+\upalpha_{\bullet}}\right).
	\end{align*}
    
 The values for the weight $w$ and the hyperparameter $\alpha$ are chosen through a cross-validation procedure. Firstly, we considered a grid of 20 equally spaced points in the intervals $[0,1]$ and $(0.001, 20]$ for $w$ and $\alpha$, respectively. Then, for each pair $(w_i,\alpha_i)$, $i=1, \ldots, 400$, the Brier scores of the first-half matches (190 matches) of each championship was computed. The pair of values $(w^*,\alpha^*)$ which provided the smallest score was then chosen to predict the matches of the second half of the same championship. Before this was done, however, the counts of each team were used to update the prior $\mathcal{D}(\alpha^*,\alpha^*,\alpha^*)$ in the same manner as described in Section \ref{sec::Mn_Dir1}.

Table \ref{tab::optimalValues} displays the optimal values of $\alpha$ and $w$ chosen for each championship. 
Note that the values are generally not far from those used in the model $Mn-Dir_1$, $\alpha=1$ and $w=1/2$.
        
\begin{table}[ht]
\centering
		\caption{Optimal values of $\alpha$
        and $w$ for each year in the model $Mn-Dir_2$.}
\begin{tabular}{rrr}
  \hline
 Year & $\alpha^*$ & $w^*$ \\ 
  \hline
 2006 & 3.16 & 0.53 \\ 
 2007 & 2.63 & 0.63 \\ 
 2008 & 1.05 & 0.42 \\ 
 2009 & 2.63 & 0.42 \\ 
 2010 & 2.11 & 0.58 \\ 
 2011 & 2.11 & 0.53 \\ 
 2012 & 1.58 & 0.53 \\ 
2013 & 2.63 & 0.79 \\ 
2014 & 3.16 & 0.63 \\ 
   \hline
\end{tabular}
\label{tab::optimalValues}
\end{table}

	One may argue that, in this case, data is being used twice in the same model---in the same spirit of empirical Bayes models---and therefore that the computation of weights is arbitrary.
	Even though these critiques are well founded, we believe that every choice to compute weights would be arbitrary.
	Ours was based on plain empirical experience, nothing more.

	\section{Results}
	\label{sec::results}
	
	Brazilian football championships are disputed by 20~teams that play against each other twice (home and away) and the team with more points after all matches are played is declared champion.
	Therefore, 380 matches are played per championship, 190 in each half.
	The last four teams are relegated to a minor division and the first four play Copa Libertadores (South America champions league).
	Our analysis comprised the championships from 2006 to 2014, because it was only in 2006 that this form of dispute was implemented in the Brazilian national championships. 
    The predictions for the models Arruda, Lee, Bradley-Terry and the proposed multinomial ones Mn-Dir1 and Mn-Dir2 were assessed according to their accuracy and calibration. The accuracy of the predictions was measured using different scoring rules (Brier, Spherical and Logarithmic) and also the proportion of errors. For an explanation of the scoring rules, the proportion of errors and calibration see the Appendix.

	As explained above, the Arruda model uses results of the previous twelve months to predict future matches, but we have no information about how this is done.
	This fact puts the Arruda model on a privileged position at the beginning of each championship.
	Hence, trying to put all the models on equal footing, we used the first-half matches to estimate the Lee and Bradley-Terry models, and as prior information for the multinomial-Dirichlet models as described in Sections \ref{sec::Mn_Dir1} and \ref{sec::Mn_Dir2}.
	Thus, the models were compared using only the predictions for matches of the second half, i.e. we effectively scored the predictions made for 1710 matches (190 matches of nine championships).
The Lee and Bradley-Terry models were fitted using the software R and the multinomial-Dirichlet models were fitted using Python. See \cite{r} and \cite{python}.

\begin{figure}[H]
		\centering
		\includegraphics[page=1,scale=0.3]{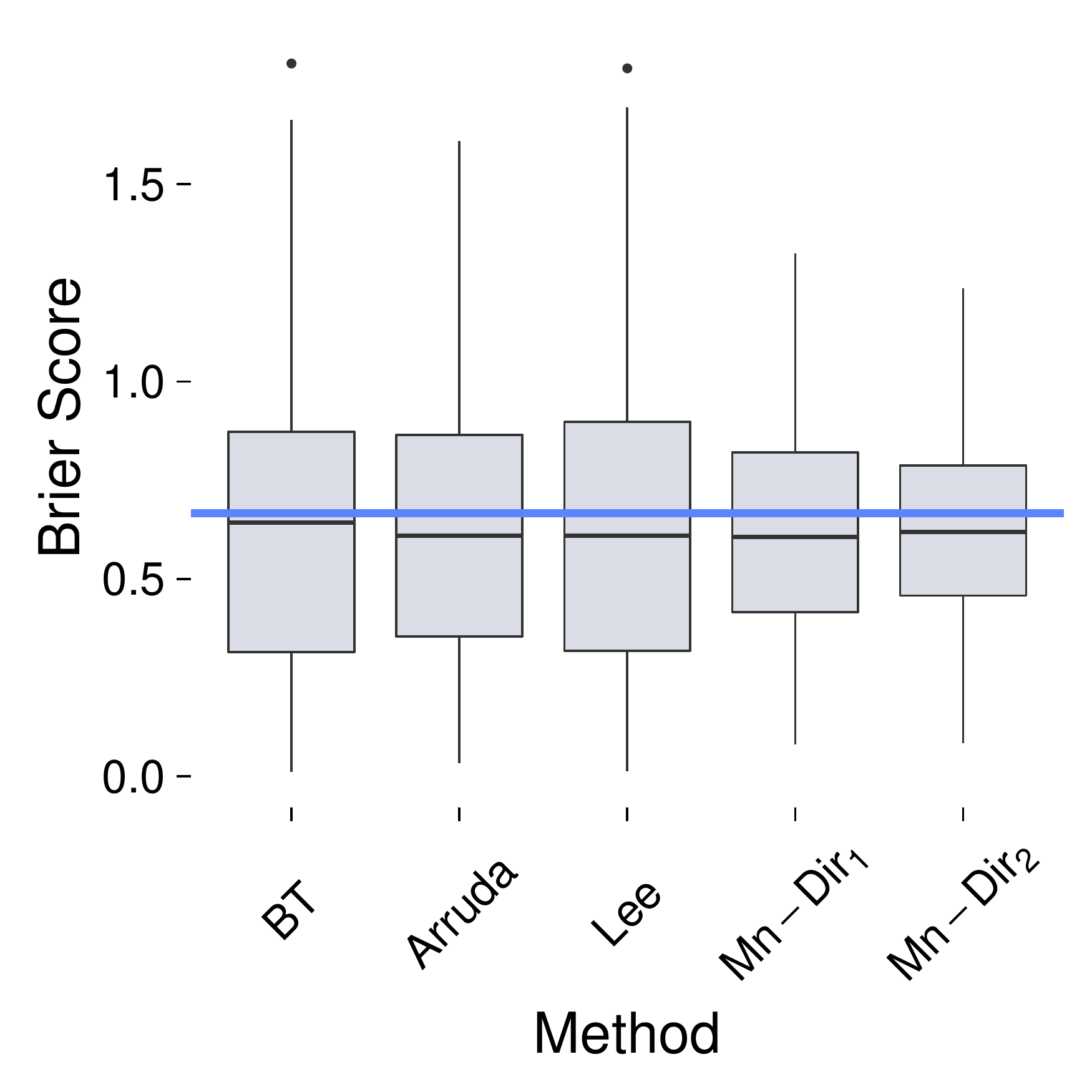}
		\includegraphics[page=2,scale=0.3]{futebolComparacaoModelosForPaperReview.pdf}\\
		\includegraphics[page=3,scale=0.3]{futebolComparacaoModelosForPaperReview.pdf}
		\includegraphics[page=4,scale=0.3]{futebolComparacaoModelosForPaperReview.pdf}
		\caption{Scores and proportion of errors of the various predictive methods. Horizontal line represents the score of the trivial prediction $(1/3,1/3,1/3)$.}
		\label{fig::scores}
	\end{figure}

\begin{figure}[H]
		\centering
        \begin{subfigure}{0.88\linewidth}
		\includegraphics[page=1,scale=0.45]{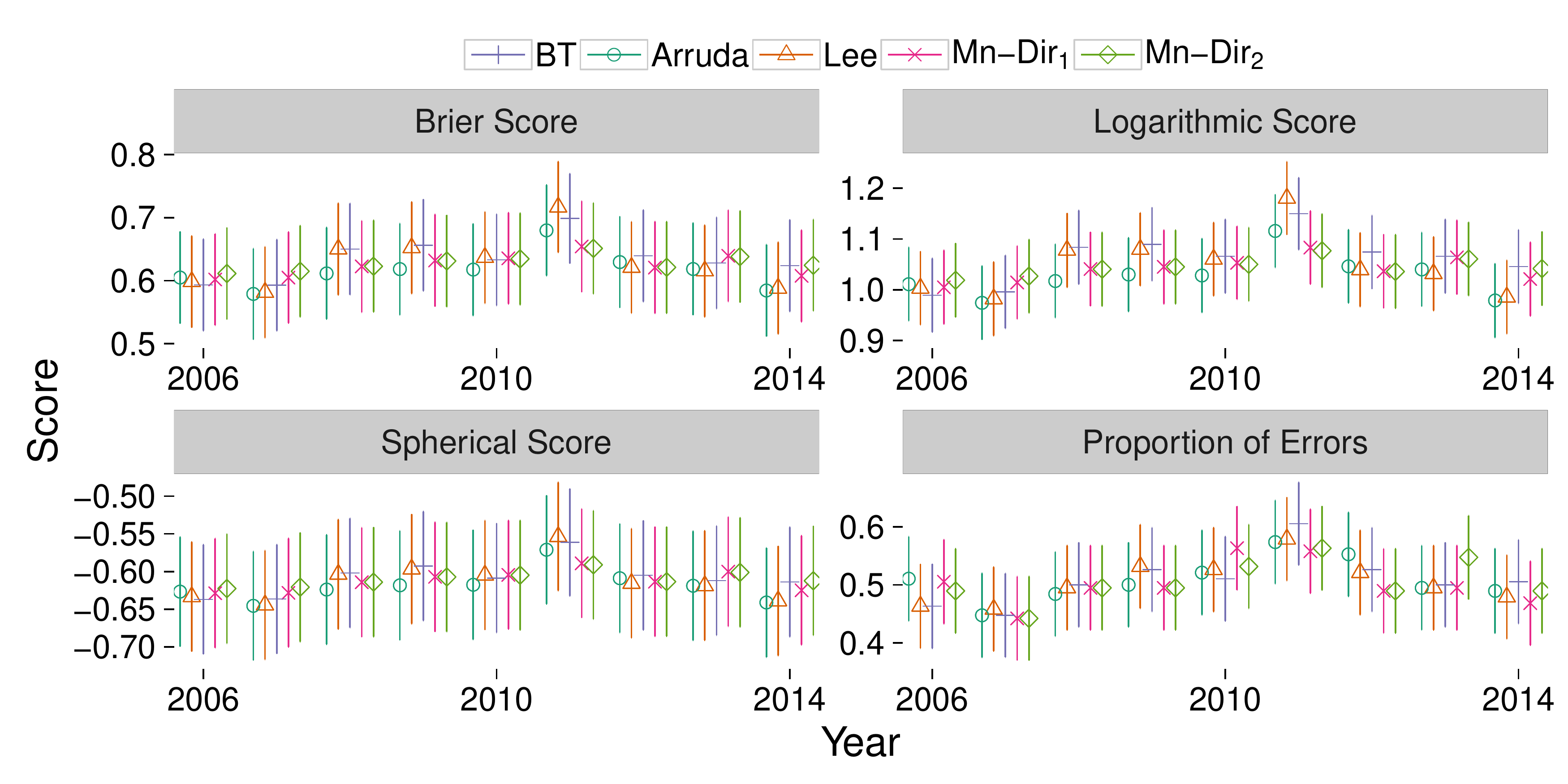}
        \caption{}
        \end{subfigure}\\
        \begin{subfigure}{0.55\linewidth}
      	\includegraphics[scale=0.35]{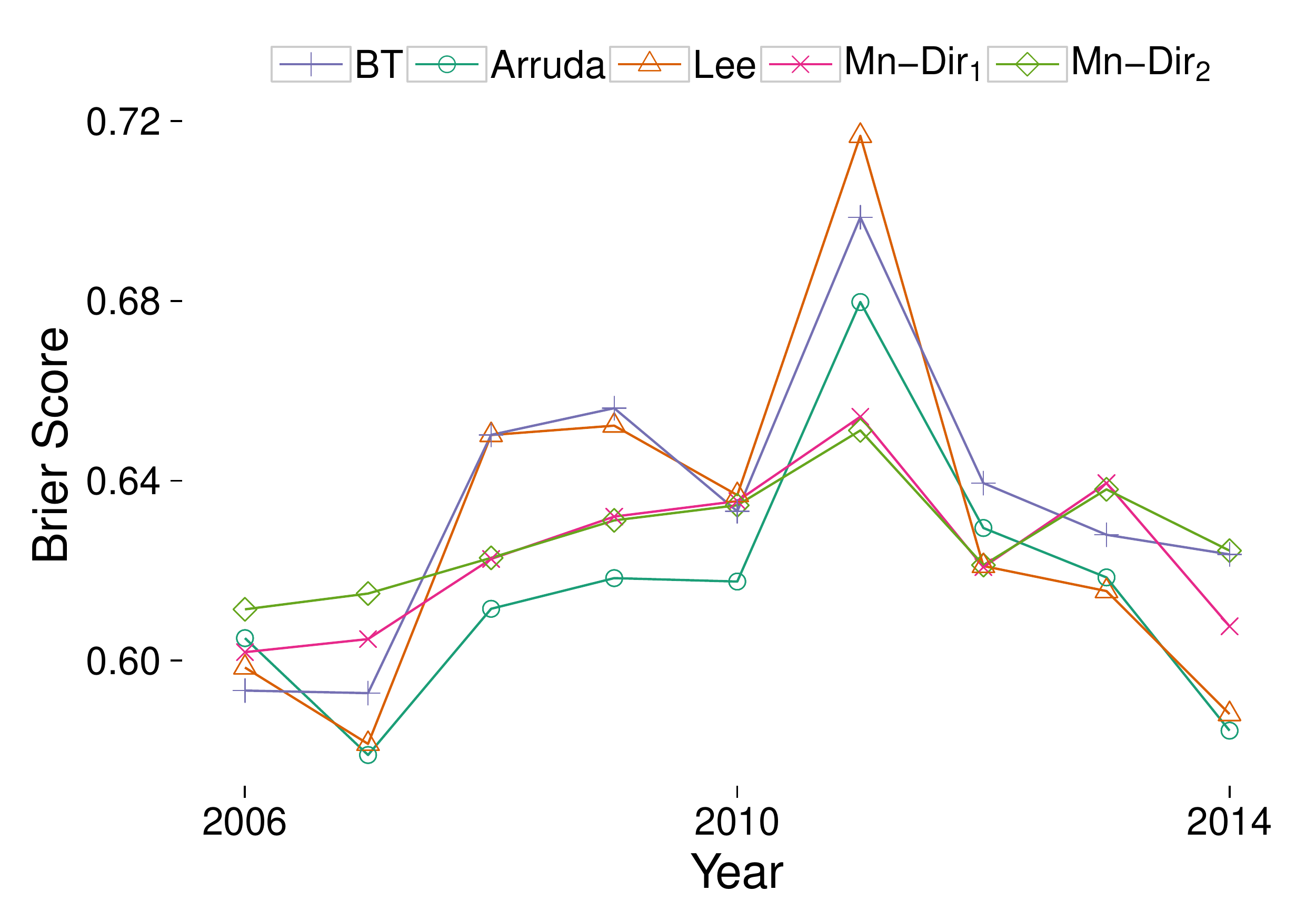}
                        \caption{}
 \end{subfigure}
		\caption{Means and standard errors of each measure of performance by year. Plot (b) shows the same information for the Brier scores, but without standard errors.}
		\label{fig::scoresYear}
	\end{figure}

	Figure~\ref{fig::scores} displays the box plots of the scores and proportion of errors of the six models in study (the lower the score, the more accurate the prediction). According to all scoring rules, all methods presented similar performance, and they were more accurate than the trivial prediction $(1/3,1/3,1/3)$, displayed in the plots
	as an horizontal line.	Using the mean scores and their standard errors displayed in Table \ref{tab::brier}, one can see that none of the 95\% confidence intervals for the mean score contained the score given by the trivial prediction
	(0.67 for the Brier score, 1.10 for the logarithmic score, and -0.58 for the spherical score). 
	Figure \ref{fig::scoresYear} shows how the scores varied year by year in average.
	This figure also indicates that all models yielded similar results.
	
	 \begin{table}[H]
        \begin{center}
                    \caption{Mean and total scores and their standard errors for the 1710 matches.}
            \resizebox{\textwidth}{!}{%
                \begin{tabular}{cccccc}
                \hline
                Score &BT & Arruda & Lee & $Mn-Dir_1$ & $Mn-Dir_2$  \\
                                \hline
                                \hline
                & \multicolumn{5}{c}{Mean Scores} \\ \cline{2-6}
                Brier &0.635 (0.009) & 0.616 (0.007)& 0.629 (0.009)& 0.624 (0.006) & 0.628  (0.005) \\
                Spherical$\times 10$ &  -6.077 (0.062)& -6.191 (0.055)& -6.112 (0.062)& -6.123 (0.048)& -6.098 (0.041)\\
                 Logarithmic & 1.062 (0.014) & 1.027 (0.011) & 1.049 (0.013)  & 1.040 (0.009)  & 1.044 (0.008) \\
                & \multicolumn{5}{c}{Total Scores} \\ \cline{2-6}
                Brier & 1085.9 (14.9) & 1053.2 (12.6)& 1075.5 (14.9)& 1067.59 (10.4) & 1073.5  (8.8)  \\
                Spherical$\times 10$ &  -1039.2 (10.5)& -1058.7 (9.4)&  -1046.9 (10.6)& -1047.0 (8.2)& -1042.7 (7.1) \\
                 Logarithmic & 1816.4 (23.3) &  1755.7 (18.0) & 1793.6 (21.9)  & 1778.8 (15.5)  & 1785.2 (13.0)  \\
                \hline
            \end{tabular}}
            \label{tab::brier}
        \end{center}
    \end{table}

In order to formally check if all models have similar predictive power, we tested the hypotheses that all six models have the same average score.
We did this by using a repeated measures ANOVA, a statistical test that takes into account the dependency between the observations (notice that each match is evaluated by each model).
In order to perform multiple comparisons, we adjusted $p$-values so as to control the false discovery rate.
All  metrics presented significant differences at the level of significance of 5\% ($p$-value $<0.01$ in all cases, see Table \ref{tab::anova}), except for the proportion of errors, where no difference was found. Post-hoc analyses 
are displayed in Table \ref{tab::postHoc}. Along with
 Table \ref{tab::brier}, one concludes that,
for the Brier score, differences were found only
between Mn-Dir$_1$ versus BT (the former had better performance), Mn-Dir$_2$ versus Arruda (the former had worse performance), and Arruda versus Lee (the former had better performance). 
For the spherical score, post-hoc analyses showed that the differences were found in Mn-Dir$_2$ versus Arruda (the former had worse performance) and BT versus Arruda (the former had worse performance). Finally, for the logarithmic score, post-hoc analyses showed that the differences were found in Mn-Dir$_1$ versus BT (the former had better performance), Mn-Dir$_2$ versus BT (the former had better performance), Lee versus Arruda (the former had worse performance), and Mn-Dir$_2$ versus Arruda (the former had worse performance).

These results therefore indicate that while the multinomial-Dirichlet models presented similar performances, they were better than BT and comparable to Lee.
It is clear that the Arruda model presented the best performance, although the predictions from Mn-Dir$_1$ were not significantly different from it, according to all scoring rules.
Hence, while BT lead to worse predictions than its competitors, Arruda was slightly better than some of its competitors, but roughly equivalent to
Mn-Dir$_1$.

\begin{table}[H]
\centering
\caption{ANOVA comparing the performance
of all prediction models under the various scores.}
\begin{tabular}{rrrrrr}
  \hline
 Score & Factor& num. d.f. & den. d.f. & F-value & p-value \\ 
  \hline
    \hline
\multirow{2}{*}{Brier} & Intercept &   1 & 8545.00 & 8423.04 & $<$0.01$^*$ \\ 
&  Model &   4 & 6836 &  5.15  & $<$0.01$^*$ \\   \hline
\multirow{2}{*}{Spherical} & Intercept &   1 & 6836 & 14650.89  & $<$0.01$^*$ \\ 
&  Model &   4 & 6836 & 3.96 & $<$0.01$^*$ \\   \hline
\multirow{2}{*}{Logarithmic} & Intercept &   1 & 6836 & 10876.28 & $<$0.01$^*$ \\ 
&  Model &   4 & 6836 &6.76 & $<$0.01$^*$ \\   \hline
Proportion  & Intercept &   1 & 6836 & 2139.93 & $<$0.01$^*$ \\ 
of Errors &  Model &   4 & 6836 & 0.31 & 0.86 \\   \hline
   \hline
\end{tabular}
\label{tab::anova}
\end{table}

\begin{table}[H]
\centering
\footnotesize
\caption{Post-hoc analyses comparing the performance
of all prediction models under the various scores.}
\begin{tabular}{llllll}
  \hline
Score & Comparison & Estimate & Std. Error & z-value & p-value \\   \hline \hline
\multirow{10}{*}{Brier} & Arruda - BT & -0.02 &  0.00 & -4.59 & $<$0.01$^*$ \\ 
&  Lee - BT & -0.01 & 0.00 & -1.47 & 0.24 \\ 
 & Mn-Dir$_1$ - BT & -0.01 & 0.00 & -2.58 & 0.04$^*$ \\ 
 & Mn-Dir$_2$ - BT & -0.01 & 0.00 & -1.75 & 0.16 \\ 
 & Lee - Arruda & 0.01 & 0.00 & 3.12 & 0.01$^*$ \\ 
 & Mn-Dir$_1$ - Arruda & 0.01 & 0.00 & 2.01 & 0.11 \\ 
 & Mn-Dir$_2$ - Arruda & 0.01 & 0.00 & 2.84 & 0.02$^*$ \\ 
 & Mn-Dir$_1$ - Lee & -0.00 & 0.00 & -1.11 & 0.36 \\ 
 & Mn-Dir$_2$ - Lee & -0.00 & 0.00 & -0.28 & 0.79 \\ 
 & Mn-Dir$_2$ - Mn-Dir$_1$ & 0.00 & 0.00 & 0.83 & 0.48 \\    \hline
\multirow{10}{*}{Spherical}  &Arruda - BT & -0.01 & 0.00 & -3.89 & $<$0.01$^*$ \\ 
 & Lee - BT & -0.00 & 0.00 & -1.54 & 0.23 \\ 
 & Mn-Dir$_1$ - BT & -0.00 & 0.00 & -1.55 & 0.23 \\ 
  &Mn-Dir$_2$ - BT & -0.00 & 0.00 & -0.69 & 0.57 \\ 
  &Lee - Arruda & 0.01 & 0.00 & 2.35 & 0.06 \\ 
  &Mn-Dir$_1$ - Arruda & 0.01 & 0.00 & 2.34 & 0.06 \\ 
  &Mn-Dir$_2$ - Arruda & 0.01 & 0.00 & 3.20 & 0.01$^*$ \\ 
  &Mn-Dir$_1$ - Lee & -0.00 & 0.00 & -0.02 & 0.99 \\ 
  &Mn-Dir$_2$ - Lee & 0.00 & 0.00 & 0.85 & 0.59 \\ 
  &Mn-Dir$_2$ - Mn-Dir$_1$ & 0.00 & 0.00 & 0.87 & 0.59 \\ 
    \hline
\multirow{10}{*}{Logarithmic} &Arruda - BT & -0.04 & 0.01$^*$ & -5.28 & $<$0.01 \\ 
 & Lee - BT & -0.01 & 0.01 & -1.99 & 0.08 \\ 
 & Mn-Dir$_1$ - BT & -0.02 & 0.01 & -3.28 & 0.01$^*$ \\ 
 & Mn-Dir$_2$ - BT & -0.02 & 0.01 & -2.72 & 0.02$^*$ \\ 
 & Lee - Arruda & 0.02 & 0.01 & 3.29 & 0.01$^*$ \\ 
 & Mn-Dir$_1$ - Arruda & 0.01 & 0.01 & 2.01 & 0.08 \\ 
 & Mn-Dir$_2$ - Arruda & 0.02 & 0.01 & 2.57 & 0.03$^*$ \\ 
 & Mn-Dir$_1$ - Lee & -0.01 & 0.01 & -1.29 & 0.27 \\ 
 & Mn-Dir$_2$ - Lee & -0.00 & 0.01 & -0.73 & 0.54 \\ 
 & Mn-Dir$_2$ - Mn-Dir$_1$ & 0.00 & 0.01 & 0.56 & 0.59 \\ 
\hline
\end{tabular}
\label{tab::postHoc}
\end{table}

\begin{figure}[H]
\centering
		\begin{subfigure}{0.48\linewidth}       \includegraphics[page=13,scale=0.48]{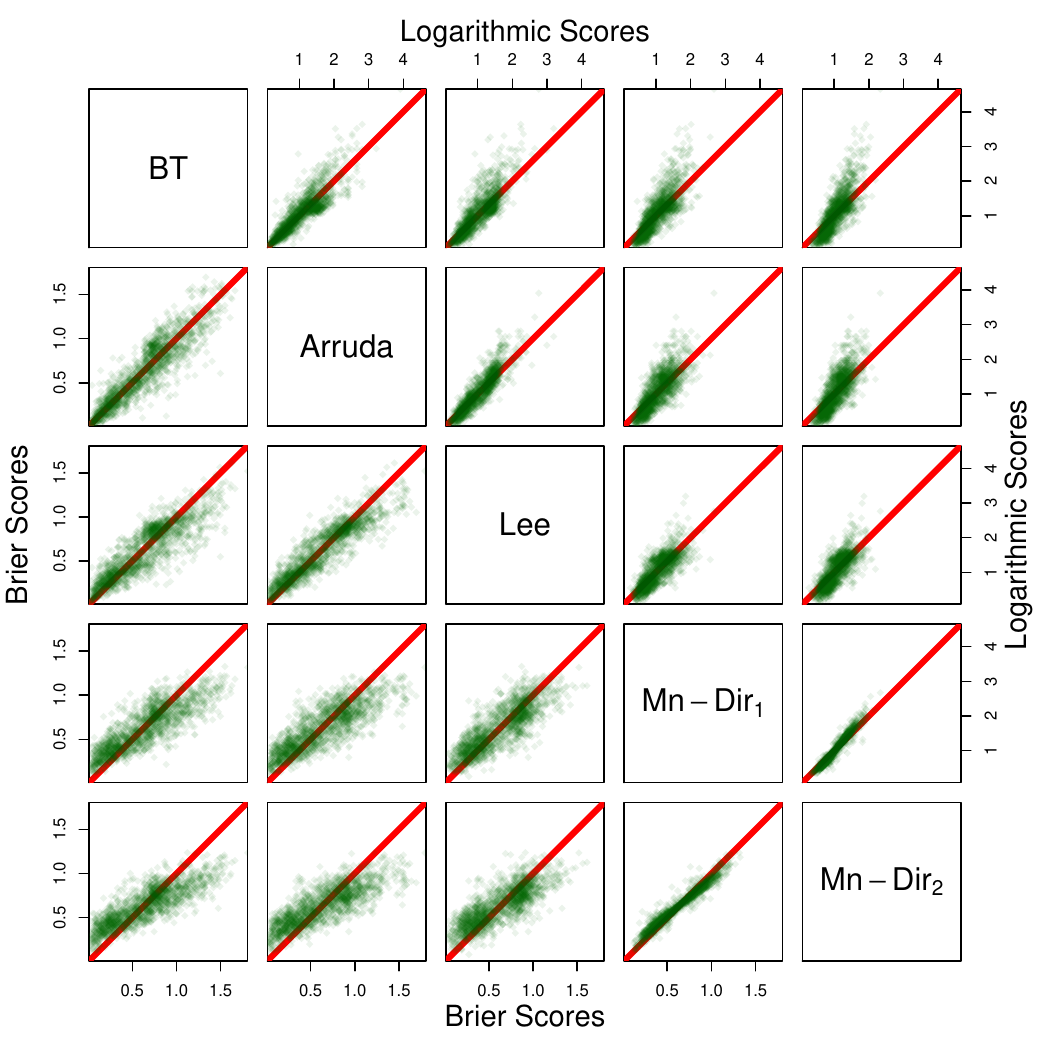}
        \caption{}
		\end{subfigure}
        \begin{subfigure}{0.48\linewidth}      \includegraphics[page=14,scale=0.48]{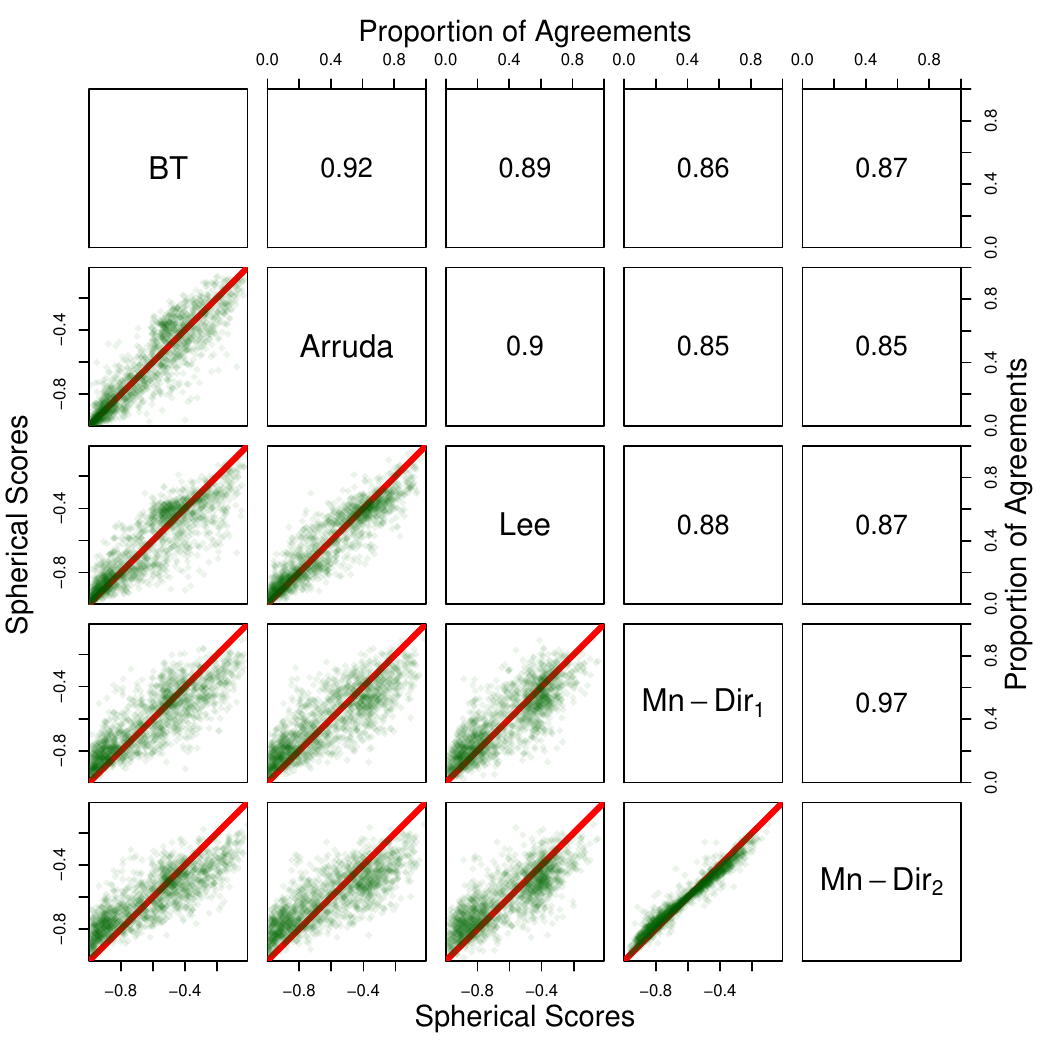}
                \caption{}
		\end{subfigure}
\caption{Pairwise comparisons of the various scores. (a): upper right plots display Logarithmic Scores;
lower left plots display Brier Scores. (b): upper right plots display proportion of agreements between methods (i.e., proportion of times the conclusions are the same; see the Appendix); lower left  plots display Spherical Scores. Lines represent} the identity $y=x$.
\label{fig::scores2}
\end{figure}

We further illustrate this point in Figure \ref{fig::scores2}, where the plots display the scores of each match for every couple of models considered. The plots show that all methods performed similarly, and that the multinomial-Dirichlet models are the ones that agreed the most.

\begin{figure}[H] \centering
	\includegraphics[page=5,scale=0.3]{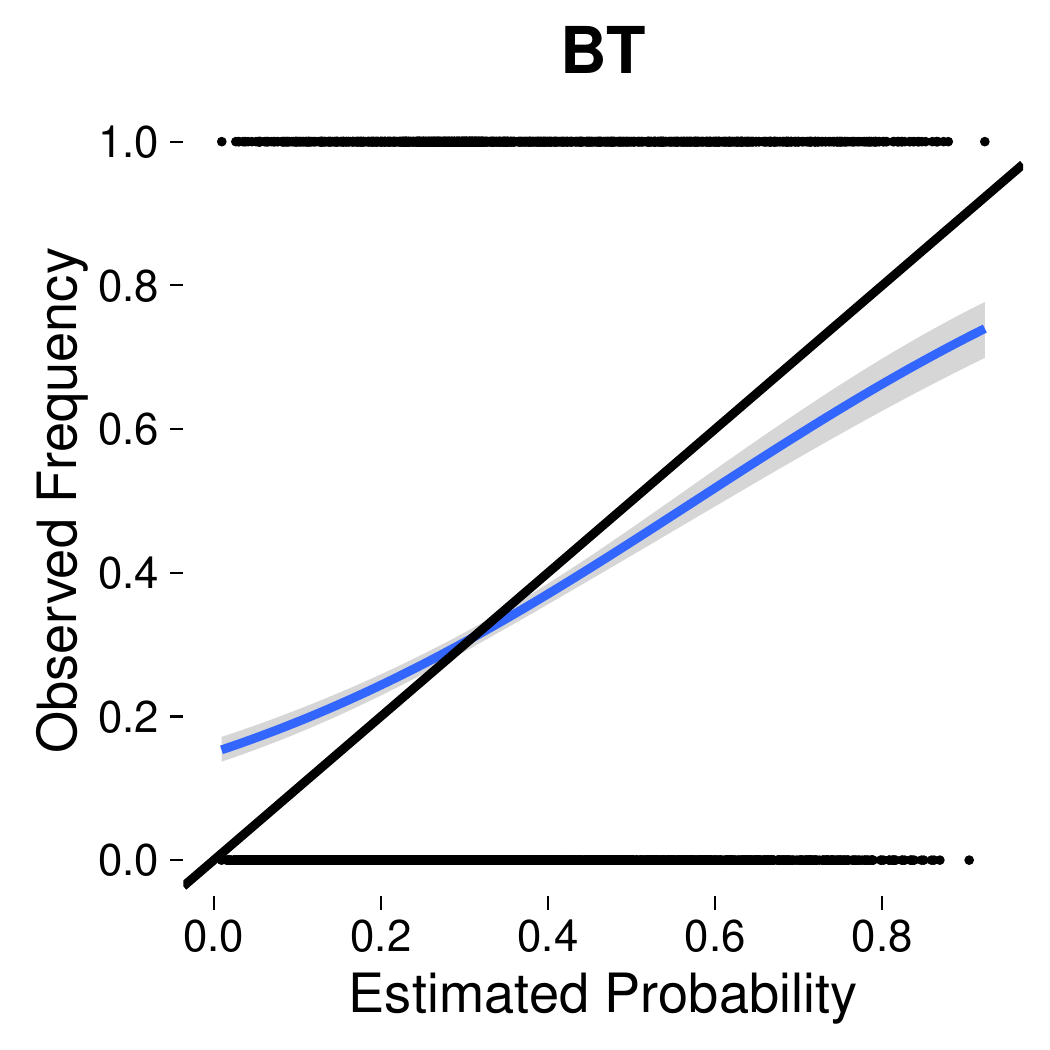}
	\includegraphics[page=6,scale=0.3]{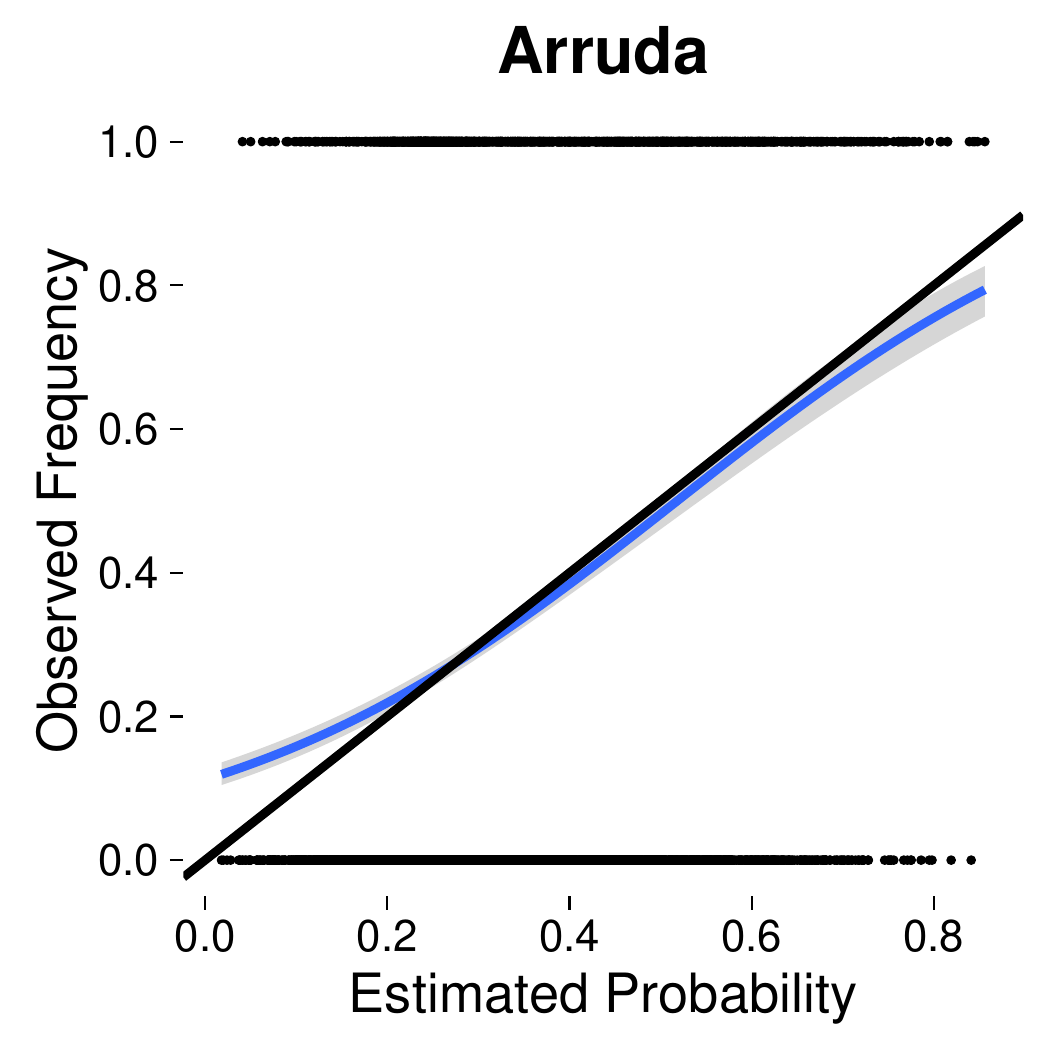}
	\includegraphics[page=7,scale=0.3]{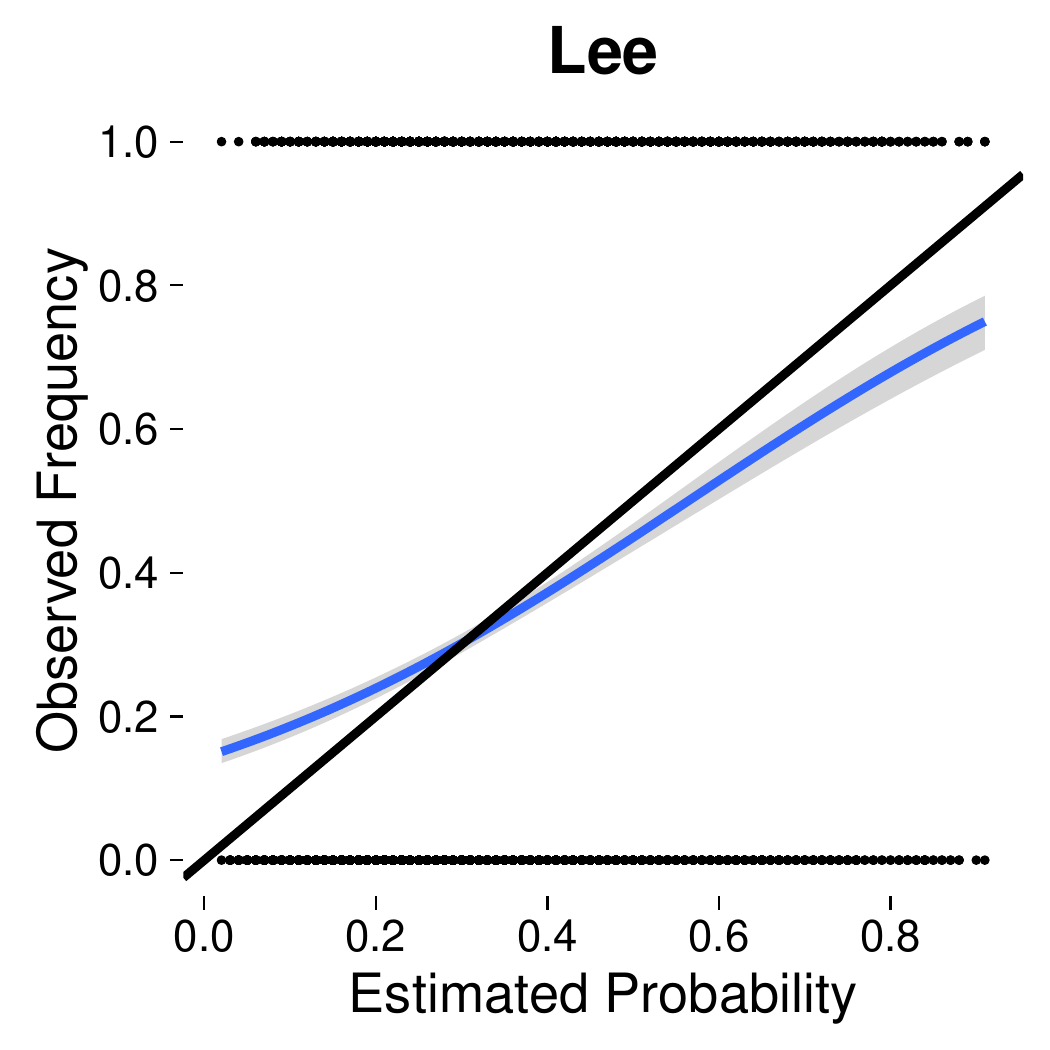}\\
	\includegraphics[page=8,scale=0.3]{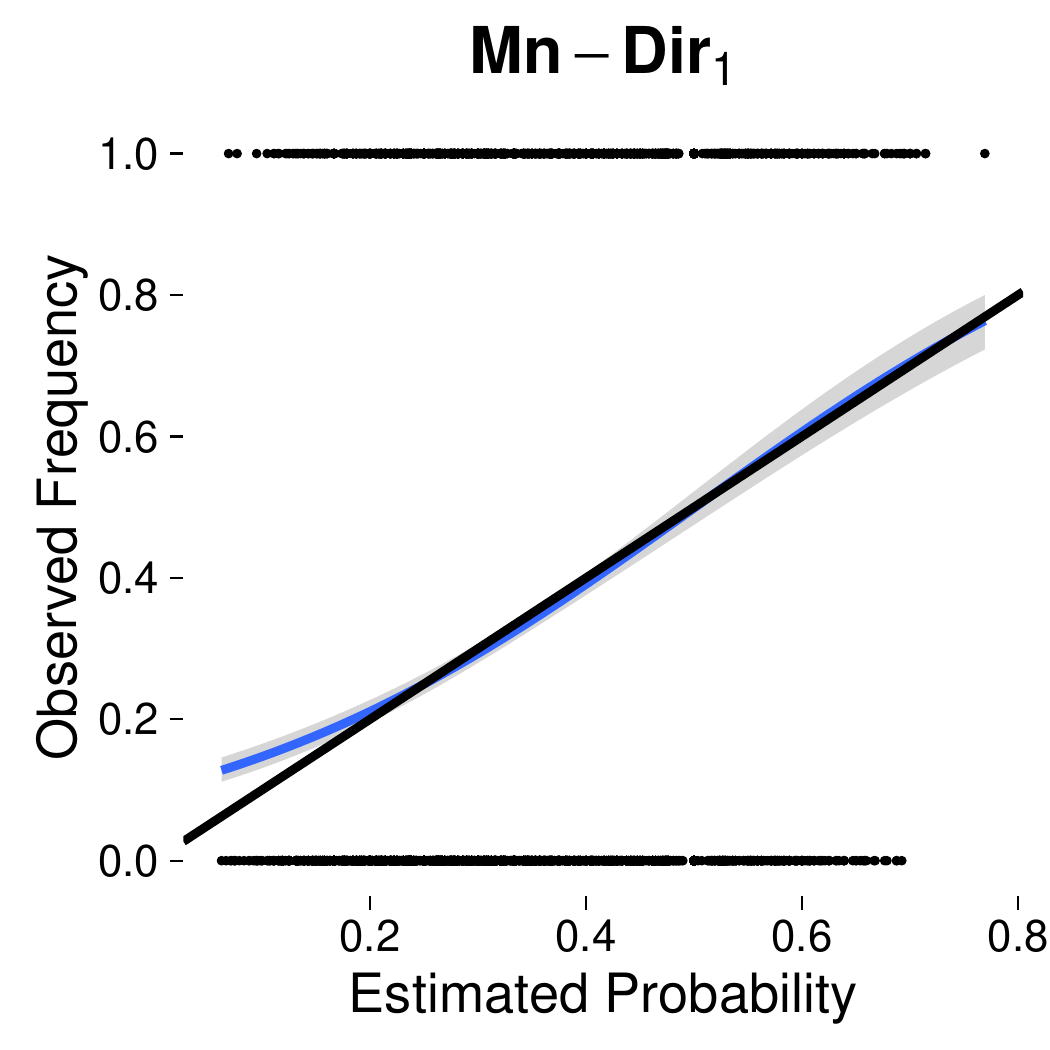}
	\includegraphics[page=9,scale=0.3]{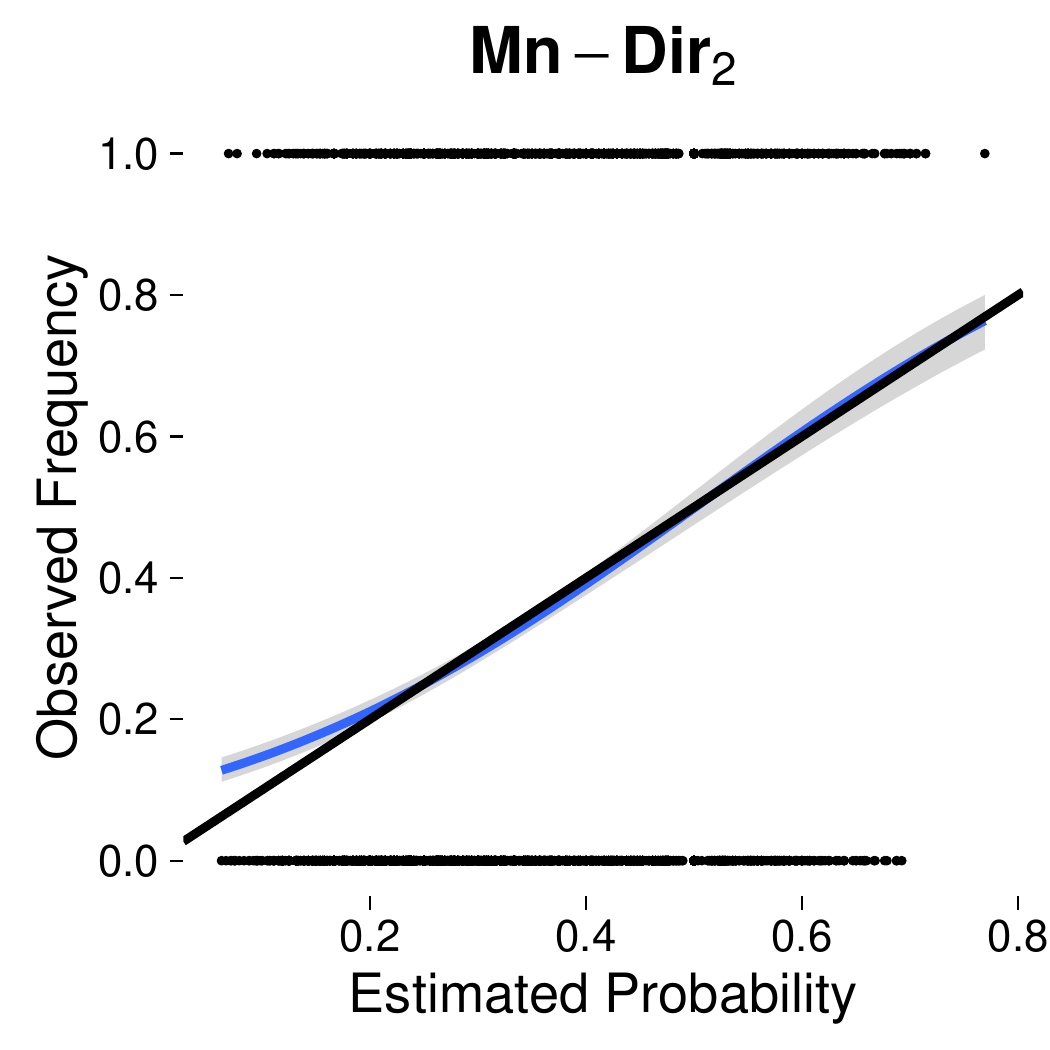}
	\caption{Calibration of the various predictive methods: estimates of
	occurrence frequency obtained by smoothing splines, with 95\%
	confidence bands. Black line is the identity $y=x$. }
	\label{fig::calibration}
\end{figure}

We also evaluated how reasonable were the predictions by assessing the calibration of the methods considered, i.e., by evaluating how often events which have assigned probability $p$ (for each $0<p<1$) happened (see the Appendix).
If these observed proportions are close to $p$, one concludes that the methods are well-calibrated. The results are displayed in Figure \ref{fig::calibration}. Because the Arruda and multinomial-Dirichlet models have curves that are close to the identity ($45^{\text{o}}$ line), we conclude that these methods are well-calibrated. On the other hand, BT and Lee seem to be poorly calibrated, over-estimating probabilities for the most frequent events.

\subsection{Goodness of fit and information measures}

We also evaluated the goodness of fit of each model by computing, for each team $t$, the following statistics:
	$$e^H_t = \sum_{i \in H_t} \widehat{p}_{t,i}\ \ \  \mbox{ and } \ \ \  e^A_t = \sum_{i \in A_t} \widehat{p}_{t,i},$$ 
	where $\widehat{p}_{t,i}$ is the estimated probability team $t$ wins the $i$-th match,
	$H_t$ is the set of matches team $t$ played as home team, and $A_t$ the set of matches team $t$ played away.
	We then computed a $\chi^2$ statistic
	$$\chi^2_o = \sum_{t} \frac{(e^H_t-o^H_t)^2}{e^H_t}+\frac{(e^A_t-o^A_t)^2}{e^A_t},$$
	where $o^H_t$ is the number of times team $t$ won playing home and $o^A_t$ is the number of times team $t$ won playing away.
	We then compared $\chi^2_o$ to a $\chi^2$
	distribution with 40 degrees of freedom (twice the number of teams of each championship).
	Since we did not fit the Arruda model, this was the only goodness of fit measure we could compute.
	
	The values of the statistics and their corresponding $p$-values are displayed in Table \ref{tab::goodness}.
	Except for the BT model, all other methods presented
	reasonable goodness of fit, in particular, the multinomial-Dirichlet model 1, which presented the smaller chi-square statistics, thus indicating better fit.

	\begin{table}[H]
		\begin{center}
			\begin{tabular}{cccccc}
				\hline
				Score &BT & Arruda & Lee & $Mn-Dir_1$ & $Mn-Dir_2$  \\
				\hline
				\hline
				$\chi^2_o$ &112.8 & 76.9& 91.5& 61.5 & 77.2 \\
				$p$-value & 0.001 & 0.48 &0.14  & 0.91  & 0.50 \\
				\hline
			\end{tabular}
			\caption{Goodness of fit statistics for the models considered.}
			\label{tab::goodness}
		\end{center}
	\end{table}

\begin{figure}[H] \centering
		\includegraphics[page=11,scale=0.4]{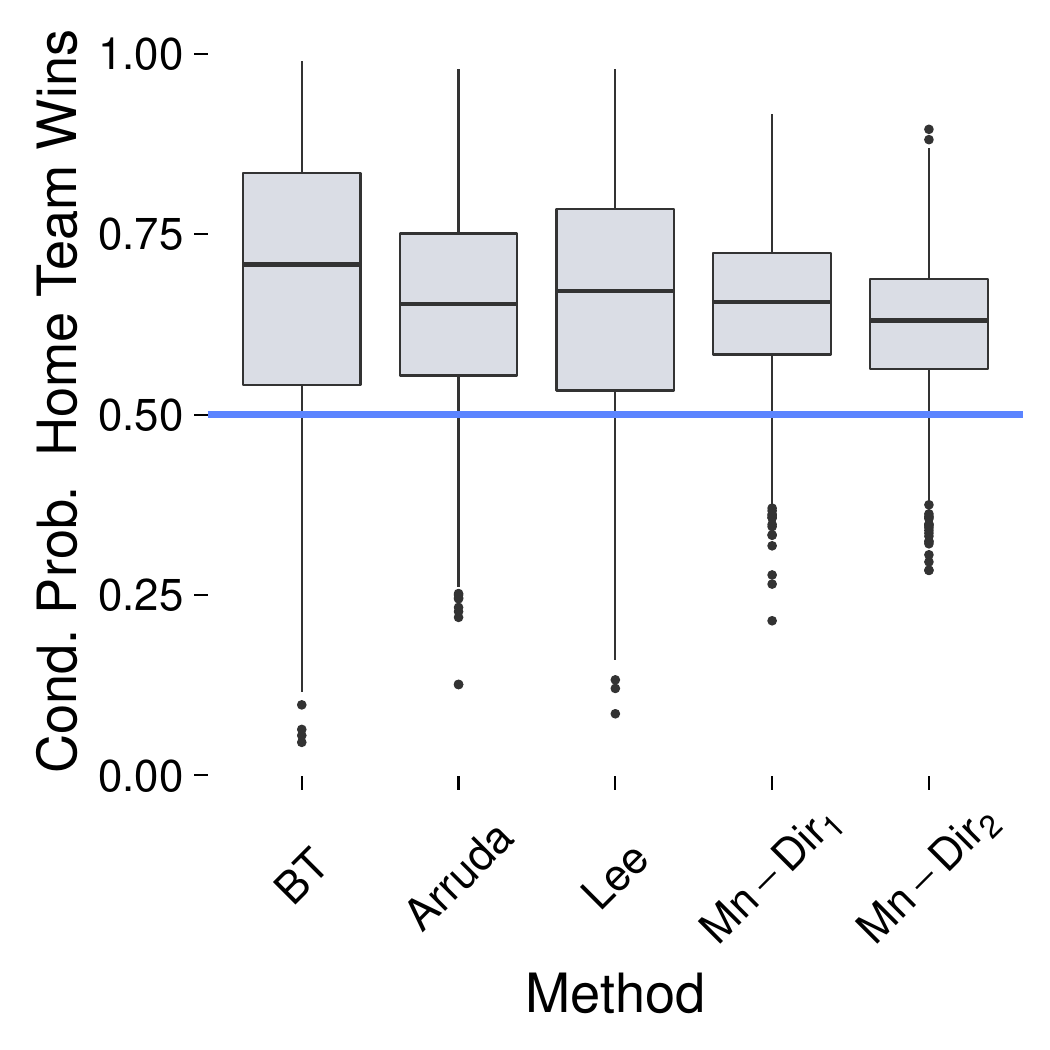}
		\caption{Conditional probability that the home team wins given there
			is no draw. Horizontal line indicates a 50\% probability.}
		\label{fig::probMandante}
	\end{figure}

To have a deeper understanding about the probabilities given by each method, Figure \ref{fig::probMandante} displays the estimated conditional probability that the home team wins assuming the match will not be a tie.
All models assigned higher probabilities to the home team, showing that they captured the well-known fact known as home advantage, peculiar in football matches and other sport competitions \citep{Pollard86, Clarke95, Nevill99}. 
	
\begin{figure}[H]
				\centering
				\includegraphics[page=10,scale=0.3]{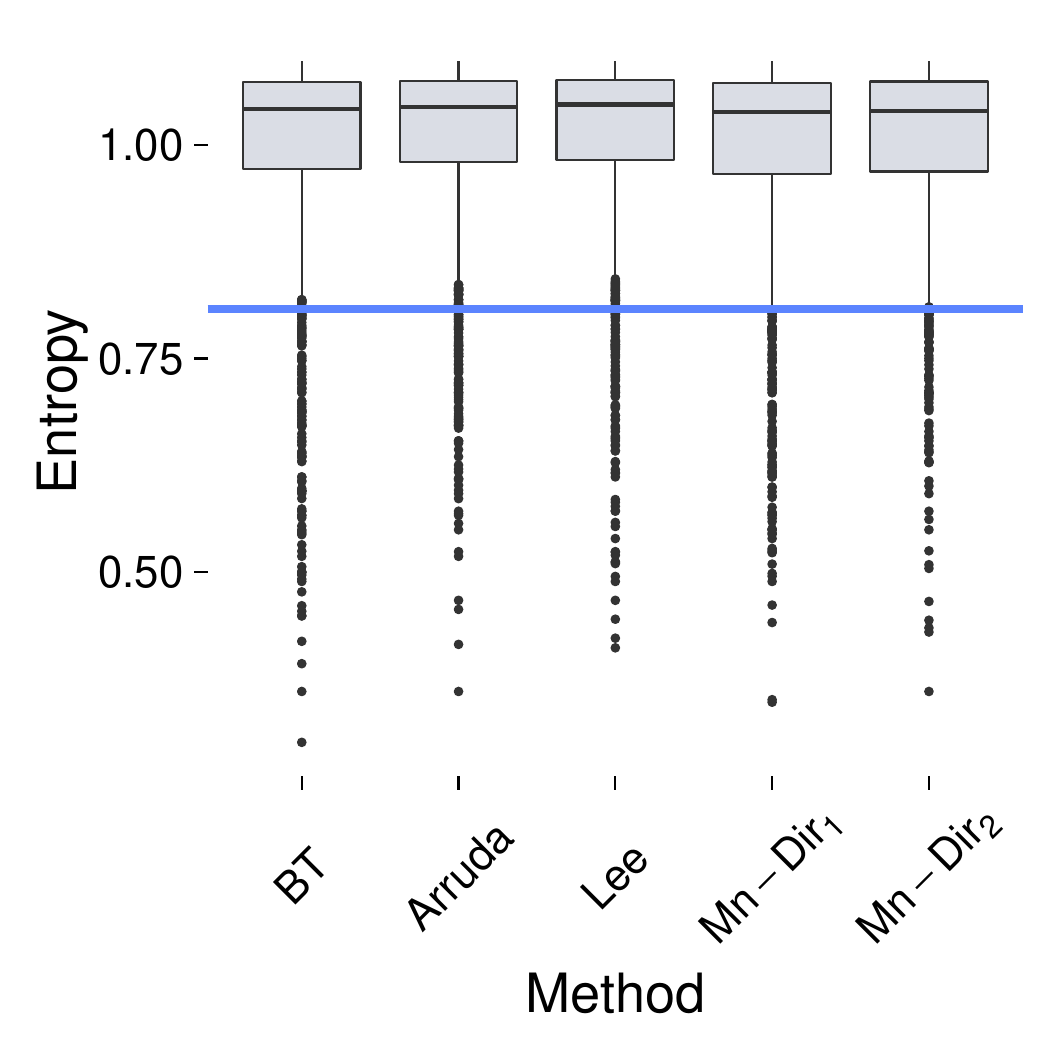}
				\caption{Entropy of the predictions of the various methods. Horizontal line represents the entropy of the trivial prediction $(1/3,1/3,1/3)$.}
				\label{fig::entropy}
			\end{figure}

In order to check how informative the predictions provided by the six models were, we computed the entropy of their predictions. 
Recall that the entropy of a prediction $(p_1,p_2,p_3)$ is given by $- \sum_{i=1}^3 p_i \log{p_i}$.
Figure \ref{fig::entropy} presents the box plots of the entropy of each prediction for all the studied models.
Since the entropy values are smaller than 1.09, the plots show that the predictions were typically more informative than the trivial prediction. 
Nevertheless, all methods yielded similar entropies on average, that is, none of them provided more informative probabilities.

Summarizing our conclusions we may say that, for the matches considered in our analysis, all the studied models yielded better predictions than the trivial prediction. 
In particular the multinomial-Dirichlet models were well-calibrated, while the Lee model was not.
Model Mn-Dir$_1$ presented the best goodness of fit statistic, while models Mn-Dir$_2$ and Arruda showed similar goodness-of-fit.
About the scoring rules, while the Bradley-Terry model yielded worse predictions than its competitors according to all metrics, the Arruda model was the best one according to the three scoring rules considered in this work, but not in every championship.
The scores of the predictions provided by the multinomial-Dirichlet models were, on average, similar to the scores of the Arruda model.

Therefore, we conclude that the
multinomial-Dirichlet models are competitive with standard approaches.

	\section{Final Remarks}
	\label{sec::remarks}

    The benchmark models used in this work were chosen because of their wide popularity among football fans in Brazil, despite the availability of several other models in the literature. Among them, we can cite those that model the match as a stochastic process evolving in time \citep{Dixon98, Volf2009, Titman2015}, those allowing for the team performance parameters to change along the season~\citep{Rue2000,Crowder2002,Owen2011,Koopman2015} and those modeling dependence between number of goals by means of bivariate count distributions \citep{Dixon97, Karlis2003, McHale2007, McHale2011}. 
Contrary to the multinomial models we proposed, some of these approaches are able to answer several questions, for instance, they can estimate teams' performance parameters allowing to rank the teams according to their offensive and defensive qualities, and can also predict the number of goals scored in a particular match. 

Another critical difference between the benchmark and the multinomial models is that the latter are Bayesian, while the former are not. Not only the way they use past information is different (because of frequentist and Bayesian paradigms), but also the pieces of information used in the analysis (for example, the Arruda model uses results of the previous twelve months, including other championships, while multinomial models use only the previous matches of the current championship). One can argue that this may lead to unfair comparisons, which is true if the interest is on the inferential properties of the models; our interest, however, is on prediction only, the true purpose of all of these models. For prediction comparions, there is not even a need for a probabilistic model, as we have seen in the comparisons with the trivial prediction.

Nonetheless, when we are interested only on predicting the final outcome of future matches, the multinomial-Dirichlet models can perform equally well as their complex counterparts.
The advantage of the first proposed model is that its predictions are obtained through simple calculations, without requiring numerical optimization procedures.
	The importance of such finding is directly related to the models usability in practice: professionals that use the mentioned benchmark models often say that a difficulty they face is that predictions may generate anger in football fans, which is then translated into distrust in subsequent predictions. 
    Because of the complexity of some models, they find hard to explain to the lay user how the outcomes were predicted.
	This is where using simple models pays off: the first multinomial model yields results that are easy to explain because they only involve counts of losses, wins and draws, allowing one to offer simple explanations to football fans and journalists about the proposed predictions.
	
Our work also poses several questions about probabilistic prediction of sport events.
		In particular, based on the fact that the models have similar predictive power on average, one may ask: Is there an irreducible ``randomness'' or ``degree of unpredictability'' implicit in these events?
		Is this degree an indicator of how tight or leveled is the championship being studied?
		
		A suggestion of future research is to answer these questions by considering more championships and models, and by comparing them using other scoring rules.
		We would also like to test other weighting methods in models $Mn-Dir_1$ and $Mn-Dir_2$ here proposed, and to evaluate their impact on the predictive power of the resulting predictions.
		
        Another possible extension is to explore different prior distributions for the multinomial-Dirichlet models. 
        In this work, we have predicted the second half of the season using a prior based on the first half. However, one can refine prior construction in order to enable first-half predictions. For instance, one can construct priors based on pre-season odds---e.g. odds for winning the championship, finishing in a given position--- or on rankings of the teams---such as Elo rankings---provided by experts before the beginning of each championship and this is equivalent, one may say, to use the results of previous matches from a given time span. 
    

	\section*{Appendix: scoring rules and calibration}
	\label{sec::scoring}
	
	In this appendix we describe the scoring rules, how we computed the proportion of errors and the calibration measure used in the paper.
	Firstly we provide a definition of proper scoring rules with simple examples to illustrate some of them and afterwards we describe the criterion developed to verify if the models are calibrated.

	One way to fairly rank predictive models is by using proper scoring rules, where
	the score may be interpreted as a numerical measure of how inaccurate a given probabilistic prediction was.
	
\begin{figure}[H]
		\begin{center}
			\begin{tikzpicture}[scale=2,tdplot_main_coords,axis/.style={->},thick]
			
			\draw[axis] (0, 0, 0) -- (1.4, 0, 0) node [right] {$p_1$};
			\draw[axis] (0, 0, 0) -- (0, 1.4, 0) node [above] {$p_2$};
			\draw[axis] (0, 0, 0) -- (0, 0, 1.4) node [above] {$p_3$};
			
			\coordinate  (d1) at (1,0,0){};
			\coordinate  (d2) at (0,1,0){};
			\coordinate  (d3) at (0,0,1){};
			
			\fill[gray!80,opacity=0.2] (d1) -- (d2) -- (d3)-- cycle;
			
			\draw[-, gray, thick] (0,0,1) -- (1,0,0);
			\draw[-, gray, thick] (0,0,1) -- (0,1,0);
			\draw[-, gray ,thick] (1,0,0) -- (0,1,0);
			
			\node[fill,circle,inner sep=1.5pt,label={left:$(1,0,0)$}] at (d1) {};
			\node[fill,circle,inner sep=1.5pt,label={south east:$(0,1,0)$}] at (d2) {};
			\node[fill,circle,inner sep=1.5pt,label={left:$(0,0,1)$}] at (d3) {};
			
			\draw[-latex,thick](d3) to [out=60,in=180] (1,1,2);
			
			\node[label={right:away team wins}] at (1,1,2) {};
			
			\draw[-latex,thick](d1) to [out=-90,in=180] (1,1,-1);
			
			\node[label={right:home team wins}] at (1,1,-1) {};
			
			\draw[-latex,thick](d2) to [out=-120,in=180] (1,1,-.25);
			
			\node[label={right:draw}] at (1,1,-.25) {};
			
			\node[fill, black, circle,inner sep=1.5pt] at (0.25,0.35,0.40) {};
			
			\draw[-latex,thick](0.25,0.35,0.40) to [out=90,in=180] (1,1,1.2);
			
			\node[label={right:$(0.25,0.35,0.40)$: prediction}] at (1,.9,1.2) {};
			
			\end{tikzpicture}
			\caption{Bi-dimensional simplex (gray surface): area of possible forecasts.}\label{fig:simplex}
		\end{center}
	\end{figure}
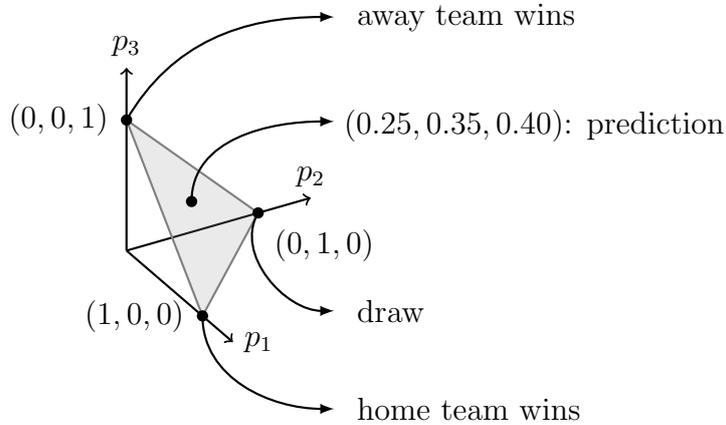

	Formally, let $X$ be a random variable
	taking values in $\mathcal{X}=\{1,2,3\}$ indicating
	the outcome of the match, with 1 standing for home win, 2 for draw and 3 for away win.  Moreover, let $P=(P_1,P_2,P_3)$ denote one's probabilistic prediction
	about $X$, \emph{i.e.}, $P$ lies in the 2-simplex set $\Delta_2=\{(p_1,p_2,p_3):p_1+p_2+p_3=1, \ p_1,p_2,p_3\geq0\}$ (see Figure
	\ref{fig:simplex}).
	A scoring rule is a function
	that assigns a real number (score) $S(x,P)$ to each $x \in \mathcal{X}$
	and $P \in \Delta_2$
	such that
	for any given $x$ in $\mathcal{X}$, the score  $S(x,P)$ is minimized when $P$ is
	$(1,0,0)$, $(0,1,0)$ or $(0,0,1)$ depending if $x$ is 1, 2 or 3, respectively.
	The score $S(x,P)$ can be thought as
	a penalty to be paid when one assigns the
	probabilistic prediction $P$ and outcome
	$x$ occurs. Also, the ``best'' possible score (\emph{i.e.}, the smallest score value) is achieved when the probabilistic prediction for the outcome of the game is perfect. A scoring rule may also be defined to be such that a large value of the score indicates good forecasts.
	
	Although many functions can satisfy the above scoring rule definition, not all of them encourage honesty and accuracy when assigning a prediction to an event. Those that do enable a fair probabilistic assignment are named \emph{proper scoring rules} \citep{lad}, which we describe in the sequence.
	
	Consider a probabilistic prediction $P^*=(P_1^*,P_2^*,P_3^*)$ for $X$.
	A proper scoring rule $S$ is a scoring rule  such that the mean score value
	$$E_{P^*}[S(X,P)]=\sum_{x=1}^3 S(x,P)P^*_x$$
	is minimized when $P=P^*$.
	In other words, if one announces
	$P$ as his probabilistic prediction
	and uses $S$ as scoring rule, the lowest
	expected penalty is obtained by reporting $P^*$, the model real uncertainty about $X$.
	Thus, the use of a proper scoring rule encourages the forecaster to announce $P^*$ (the correct one)
	as his probabilistic prediction  rather than some other quantity.
	
	In what follows, we describe in detail three proper scoring rules we use to assess the considered models.
	We also recall the concept of calibration and propose a way to measure the calibration degree of each model.
		
	\subsection*{Brier Scoring Rule}
	
	Let $P=(P_1,P_2,P_3)$ be a probabilistic prediction  for $X$.
	The Brier score for a given outcome $x\in\{1,2,3\}$ is given by
	$$S(x,P)= \sum_{i=1}^3\mathbb{I}(x=i)(1- P_i)^2+\sum_{i=1}^3\mathbb{I}(x\neq i)P^2_i,$$
	where $\mathbb{I}$ is the indicator function.
	
\begin{figure}[H]
		\centering
		
				\begin{subfigure}[b]{0.48\linewidth}        
					\centering
					
					\begin{tikzpicture}[scale=4]
					\draw [thick](0,0) -- (1.1547,0) -- (0.57735,1)-- (0,0);
					
					\node[fill, black, circle,inner sep=1.5pt] at (0.63509,0.40) {};
					\node[label={below:$p$}] at (0.63509,0.40) {};
					

					\draw[dashed,thick] (0.63509,0.40) -- (0.57735,1);
					
					\draw[decorate,decoration={brace,amplitude=5pt},xshift=-1pt,yshift=0pt] (0.63509,0.40) -- (0.57735,1) node[black,midway,xshift=-0.4cm] {$d$};

					\node[label={below:Home wins}] at (0,-0.05) {};
					\node[label={below:Draw}] at (1.1547,-0.05) {};
					\node[label={above:Away wins}] at (0.57735,1.1) {};
					
					\node[label={below:$(1,0,0)$}] at (0,0.05) {};
					\node[label={below:$(0,1,0)$}] at (1.1547,0.05) {};
					\node[label={above:$(0,0,1)$}] at (0.57735,0.95) {};
					
					\draw[dashed,thick] (-.3,1) -- (0.57735,1);
					\draw[dashed,thick] (-.3,0) -- (0,0);
					\draw[decorate,decoration={brace,amplitude=10pt},xshift=-9.5pt,yshift=0pt] (0,0) -- (0,1.01) node[black,midway,xshift=-0.7cm] 			{\large$\frac{\sqrt{6}}{2}$};
					
					
					\end{tikzpicture}

					\caption{Brier score, $d$, for victory of away team}
					\label{fig:A}
				\end{subfigure}
		\begin{subfigure}[b]{0.48\linewidth}        
			\centering
			
			\begin{tikzpicture}[scale=4]
			\draw [thick](0,0) -- (1.1547,0) -- (0.57735,1)-- (0,0);
			
			\node[fill, black, circle,inner sep=1.5pt] at (0.63509,0.40) {};
			
			\draw[dashed,thick] (0.63509,0.40) -- (0.63509,0);
			\draw[dashed,thick] (0.63509,0.40) -- (0.85159,0.525);
			\draw[dashed,thick] (0.63509,0.40) -- (0.331976,0.575);
			
			\draw[thick] (0.63509,0.07) -- (0.70509,0.07) -- (0.70509,0);
			\node[fill, black, circle,inner sep=.7pt] at (0.67009,0.035) {};			
			\draw[thick] (0.7909695,0.49) -- (0.7559695,0.5506225) -- (0.8165912,0.5856225);
			\node[fill, black, circle,inner sep=.7pt] at (0.8037803,0.5378116) {};
			\draw[thick] (0.3925977,0.54) -- (0.3575977,0.4793783) -- (0.2969759,0.5143783);
			\node[fill, black, circle,inner sep=.7pt] at (0.3447868,0.5271892) {};
			
			\draw[decorate,decoration={brace,mirror,amplitude=4pt},xshift=-0.7pt,yshift=0pt] (0.63509,0.40) -- (0.63509,0) node[black,midway,xshift=-0.4cm] {$p_3$};
			\draw[decorate,decoration={brace,mirror,amplitude=5pt},yshift=-0.5pt,xshift=0.5pt] (0.63509,0.40) -- (0.85159,0.525) node[black,midway,yshift=-0.5cm,xshift=10pt] {$p_1$};
			\draw[decorate,decoration={brace,mirror, amplitude=5pt},xshift=0.1pt,yshift=0.7pt] (0.63509,0.40) -- (0.331976,0.575) node[black,midway,xshift=5pt,yshift=10pt] {$p_2$};
			
			\node[label={below:Home wins}] at (0,0) {};
			\node[label={below:Draw}] at (1.1547,0) {};
			\node[label={above:Away wins}] at (0.57735,1) {};

			\draw[dashed,thick] (-.3,1) -- (0.57735,1);
			\draw[dashed,thick] (-.3,0) -- (0,0);
			\draw[decorate,decoration={brace,amplitude=10pt},xshift=-9.5pt,yshift=0pt] (0,0) -- (0,1.01) node[black,midway,xshift=-0.6cm] {\large$1$};
			

			\end{tikzpicture}

			\caption{Normalized simplex: $p_1+p_2+p_3=1$}
			\label{fig:B}
		\end{subfigure}
		\caption{Standard \textbf{(a)} and normalized \textbf{(b)} simplexes.}
		\label{fig:norm_stand}
	\end{figure}
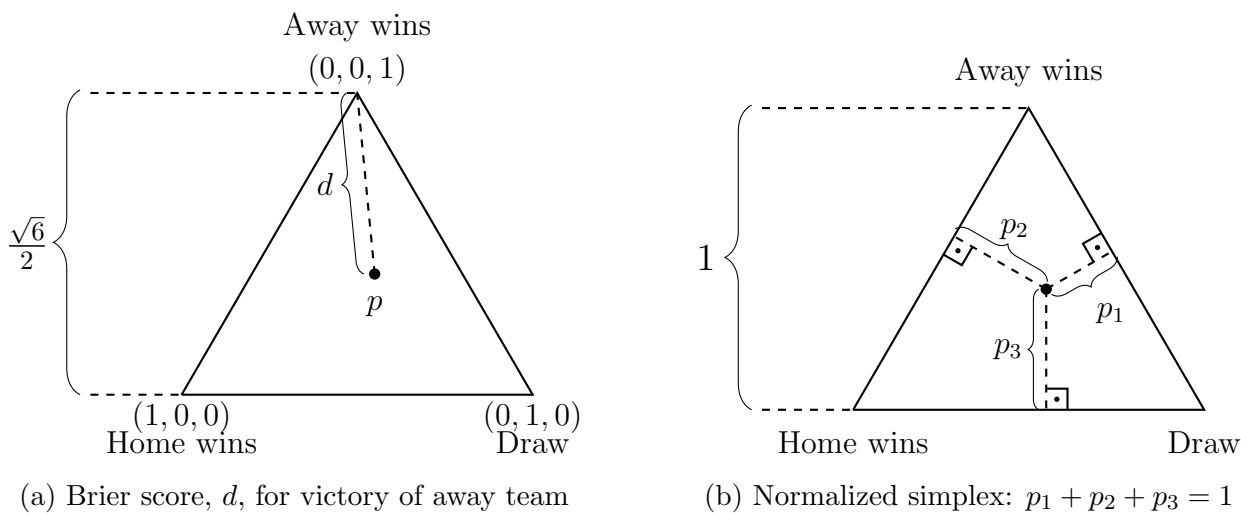

	We interpret the Brier score in the case where one of three mutually exclusive outcomes happens as in the case of a football match.
	The green surface in Figure \ref{fig:simplex} represents the 2-simplex, \emph{i.e.}, the set of points such that $p_1+p_2+p_3=1$ for non-negative values of $p_1$, $p_2$ and $p_3$.
	The triangle representing the simplex has sides of length $\sqrt{2}$ and its height is $\sqrt{6}/2$.
	Drawing a similar equilateral triangle with height 1 and sides $2\sqrt{3}/3$, it is possible to represent all points of the simplex.
	This new triangle is used to graphically display the forecast as an internal point because the sum of the distances of every point inside it to each side, representing the probability of each event, is always one.
	See Figure \ref{fig:norm_stand}.

	The Brier score for the probabilistic prediction $P=(0.25,0.35,0.40)$
	assuming the home team loses, is therefore given by $d^2=(0-0.25)^2+(0-0.35)^2+(1-0.40)^2=0.545$.
	On the other hand, the prediction $P=(0,0,1)$ achieves score  zero, the minimum for this rule.
	
	It is useful to consider the score of what we will call trivial prediction:
	$P=(1/3,1/3,1/3)$.
	This~assessment will produce a Brier score of $2/3$, no matter what is the final result of the match, providing, thus, a threshold that a good model should consistently beat, meaning, for the Brier score, that the scores of its predictions should be smaller than $0.667$.
	
	\subsection*{Logarithmic Scoring Rule}
	
	The logarithmic  score is given by
	$$S(x,P)=- \sum_{i=1}^3\mathbb{I}(x=i)\ln(P_i),$$
	
	\noindent
	which is the negative log likelihood of the event that occurred.
	
	The logarithmic score for the prediction
	$P=(0.25,0.35,0.40)$
	when the home team loses is therefore
	$-\ln(0.4)\approx 0.91$.
	On the other hand, the prediction $P=(0,0,1)$ achieves score zero, once again the minimum of this rule.
	Moreover, for the logarithmic score, the trivial prediction gives a score of approximately $1.098$.
	
	\subsection*{Spherical Scoring Rule}
	
	The spherical score is given by
	$$S(x,P)=- \frac{1}{\sqrt{\sum_{i=1}^3 P^2_i}}\sum_{i=1}^3\mathbb{I}(x=x_i)P_i,$$
	
	\noindent
	which is the negative likelihood of the event that occurred, normalized by the square-root of the sum of the assigned squared probabilities.
	
	The spherical score for the prediction
	$P=(0.25,0.35,0.40)$ assuming the home team loses, is given by
	$-0.4/\sqrt{0.25^2+0.35^2+0.40^2} \approx -0.68$.
	On the other hand, the prediction $P=(0,0,1)$ achieves score  $-1$ instead and, for this rule, the trivial prediction results in a score of approximately $-0.577$.
	
	\subsection*{Calibration and Proportion of Errors}
	\label{sec::calib}
	
	Besides scoring rules, there are other criteria used to assess the quality of different predictions. Here~we explore two of them.
	
	The first one is the proportion of errors made by the model or assessor. This is simply the proportion of mistakes made when considering the highest probability assessment.
	More precisely, the proportion of errors of a sequence of probabilistic predictions for $n$ games, $P^{(1)},\ldots,P^{(n)}$
	with
	$P^{(j)}=(P^{(j)}_1,P^{(j)}_2,P^{(j)}_3)$, is defined by
	$$\frac{1}{n}\sum_{j=1}^n \mathbb{I}\left(X_j \neq \arg \max_{x \in \{1,2,3\}} P^{(j)}_x\right),$$
	where $X_j$ is the outcome of the $j$-th match.
	
	The second concept we use is that of calibration \citep{Dawid}. Probability assertions are said to be well calibrated at the level of probability $p$ if the observed proportion of all propositions that are assessed with probability $p$ equals $p$.
	
	Because we typically do not have several predictions with the same assigned probability $p$, we obtain a plot by smoothing (\emph{i.e.}, regressing) the indicator function of whether a given result happened as a function of the probability assigned for that result.
	That is, we estimate the probability that an event occurs given its assigned probability.
	The smoothing was done via smoothing splines \citep{wahba}, with tuning parameters chosen by cross-validation.

		\bibliographystyle{apalike}
		

\end{document}